\documentclass[aps,twocolumn,floatfix,showpacs,intlimits]{revtex4}

\usepackage{graphicx,bm}
\usepackage{epsfig,color}
\usepackage{amsmath}
\usepackage{amssymb}
\usepackage{bbm}

\begin{document}

\newcommand{\rb}{{\bf r}}
\newcommand{\Rb}{{\bf R}}
\newcommand{\xb}{{\bf x}}
\newcommand{\yb}{{\bf y}}
\newcommand{\zb}{{\bf z}}
\newcommand{\kb}{{\bf k}}
\newcommand{\Kb}{{\bf K}}
\newcommand{\kbp}{{{\bf k}^\prime}}
\newcommand{\Kbp}{{{\bf K}^\prime}}
\newcommand{\dk}{\frac{d\kb}{(2\pi)^3}\,}
\newcommand{\dK}{\frac{d\Kb}{(2\pi)^3}\,}
\newcommand{\dkp}{\frac{d\kbp}{(2\pi)^3}\,}
\newcommand{\dKp}{\frac{d\Kbp}{(2\pi)^3}\,}
\newcommand{\eb}[1]{{\bf \hat e_{#1}}}
\newcommand{\st}[1]{\textcolor{red}{\sout{#1}}}
\newcommand{\sr}[1]{\textcolor{red}{#1}}

\title{Three Resonant Ultra-Cold Bosons: Off-Resonance Effects}
\author{Mattia Jona-Lasinio$^{(1)}$ and Ludovic Pricoupenko$^{(2)}$}
\affiliation
{$^{(1)}$ LENS and Dipartimento di Fisica, Universit\`a di Firenze\\  
Via Nello Carrara 1, 50019 Sesto Fiorentino, Italy\\
$^{(2)}$Laboratoire de Physique Th\'{e}orique de la Mati\`{e}re Condens\'{e}e,
Universit\'{e} Pierre et Marie Curie, case courier 121, 4 place Jussieu, 75252 
Paris Cedex 05, France.}
\date{\today}
\begin{abstract}
We solve a finite range two-channel model for three resonant 
identical bosons. The model provides a minimal description 
of the various magnetic Feshbach resonances in single 
species ultra-cold bosonic systems, including off-resonant 
scattering. We obtain important insights into the interpretation 
of seminal experiments: the three-body recombination rate measured 
in Sodium and the Efimov resonances observed in C{\ae}sium. This 
approach quantifies non universal effects appearing for a 
finite magnetic field detuning.
\end{abstract}
\pacs{34.50.Cx 03.65.Nk 03.75.Kk 05.30.Jp}
%

\maketitle

One of the main issues in current ultracold physics is to achieve highly correlated quantum gases. 
In these studies the magnetic Feshbach resonance is a crucial ingredient: it permits to tune the 
strength of the two-body interaction measured by the $s$-wave scattering length to an arbitrarily 
large value, while the atomic density remains constant. However for dilute bosonic gases, 
such highly correlated states are very unstable as a consequence of three-body recombinations 
into deep molecular bound states \cite{Ketterle_Feshbach,Stenger}. Therefore, in the last 
decade an impressive effort has been made both theoretically and experimentally, 
for a deep understanding of three-body properties in these systems. Universality concepts 
borrowed from Nuclear physics together with the specificity of ultracold atoms  where an explicit 
energy scale separation occurs in scattering processes have led to unsuspected physical insights 
in this domain \cite{Efimov,Braaten}. For example, the so-called Efimov states have been 
observed for the first time in a resonant ultracold C{\ae}sium gas \cite{Kraemer}. For scattering lengths much larger 
than the range of inter-atomic forces, Universal Theory \cite{Braaten} permits a powerful 
analysis of three-body properties without any knowledge of the short range details of the real
interatomic forces. For narrow resonances like the one observed in Ref.\cite{Stenger}, it is
possible to fully determine the Efimov spectrum at resonance using the effective range approach 
\cite{Petrov3B,Mora}. Despite their success, these approaches are not designed for a description 
of three-body properties at finite detuning, where off-resonant scattering effects come into play. 
Experiments on Efimov states in Ref.~\cite{Knoop} clearly exhibit some non universal 
behavior which cannot be taken into account by the Universal Theory. Moreover, the peak in
three-body losses in Ref.~\cite{Stenger} occurs for relatively large magnetic detuning 
where the effective range approach is inoperant. In this letter, we solve the three-boson 
problem by using a two-channel model including the short range character of interatomic forces. 
The model describes all the various type of magnetic Feshbach resonances: broad, narrow or 
in the neighborhood of a shape resonance. Results of the model compare quantitatively with 
experiments of Ref.~\cite{Stenger}, highlighting the importance of off-resonant effects 
in narrow resonances. Concerning the experiments in Refs.~\cite{Kraemer,Knoop}, our results 
show that the observed violation of universality follows from the fact that the resonance 
in the atom-dimer scattering  appears in a domain where the range of interatomic forces
is not negligible as compared to the scattering length.

We first introduce the finite range two-channel model. The two channels refer to the 
'open' channel populated by atoms which are identical ultra-cold bosons of mass $m$, 
and to the 'closed' channel where fundamentals entities are couples of two tightly 
bound atoms that we here call 'molecules'. The model takes a simple form 
in the second quantized form, where the operator $a_\kb$ annihilates an atom in the 
open channel with momentum $\kb$, while $b_\kb$ annihilates a molecule of wavevector 
$\kb$ in the closed channel. Both $a_\kb$ and $b_\kb$ obey standard bosonic 
commutation rules 
${[a_\kb,a_{\kbp}^\dagger]={[b_\kb, b_{\kbp}^\dagger]}=(2\pi)^3 \delta(\kb-\kbp)}$, 
corresponding to the choice  ${\langle \rb|\kb \rangle = \exp({i \kb \cdot \rb})}$ 
for the plane wave state. Any other commutator vanishes. The Hamiltonian is 
similar to the ones introduced for two-component or fully polarized fermions in 
Refs.~\cite{2channel,3body_pwave}:
\begin{multline}
H = \int \frac{d{\mathbf k}}{(2\pi)^3} \left[ \epsilon_{\mathbf{k}}  a^\dagger_\kb a_\kb + 
\left( \frac{ \epsilon_{\mathbf{k}}}{2} + E_{\mathrm{mol}} \right) b^\dagger_\kb b_\kb  \right] 
\\
+ \frac{g_0}{2} \int \frac{d{\mathbf k}d{\mathbf K}d{\mathbf k}'}{(2\pi)^9} 
\chi^*_\kb \chi_\kbp \, 
a^\dagger_{\frac{\Kb}{2} -\kbp} a^\dagger_{\frac{\Kb}{2}+\kbp} a_{\frac{\Kb}{2} +\kb}a_{\frac{\Kb}{2}-\kb}
\\
+ \Lambda \int \frac{d{\mathbf k}d{\mathbf K}}{(2\pi)^6}  
\left( 
\chi^*_{\bf k} \, b^\dagger_{\bf K} a_{\frac{\Kb}{2}-\kb} a_{\frac{\Kb}{2}+\kb}   + {\rm h.c.} 
\right) .
\label{eq:H}
\end{multline}
The first two terms in Eq.~\eqref{eq:H} are the kinetic operators in the open and closed 
channel respectively: ${\epsilon_{\mathbf{k}}=\hbar^2 k^2/(2m)}$, and ${E_{\rm mol}}$ 
is the internal energy of the molecular state in the closed channel, defined with respect to 
the zero energy in the open channel.  The magnetic tunability of the interaction strength 
is due to the fact that ${E_{\rm mol}}$ is an affine function of the magnetic field ${\mathcal B}$ with 
the slope ${\delta \mu}$, where ${\delta \mu}$ is the difference between the magnetic moments 
for an atomic pair in the open and closed channel. The second line in Eq.~\eqref{eq:H} mimics 
the interatomic force in the open channel which is responsible for the background scattering. 
In a real system it is characterized by an attractive van der  Waals tail, and the van der Waals 
constant $C_6$ sets up the short range scale of this pairwise interaction: 
${R_{\rm vdW} = \frac{1}{2}(m C_6/\hbar^2)^{1/4}}$. In the present model, for simplicity 
we choose a separable interaction, with a Gaussian 
weight $\chi_\kb= \exp(-k^2b^2/2)$ imposing a cut-off at short distances: 
$b \equiv O(R_{\rm vdW})$. The last term  in Eq.~\eqref{eq:H} describes the coupling 
between the two channels and  models the Feshbach resonance mechanism 
(we choose ${\Lambda \in {\mathbb R}}$). The inter-channel coupling is also 
taken into account \emph{via} the same Gaussian weight $\chi_\kb$ as in the 
open channel interacting term.

\begin{figure}[h] 
\resizebox{8 cm}{!}
{
\includegraphics{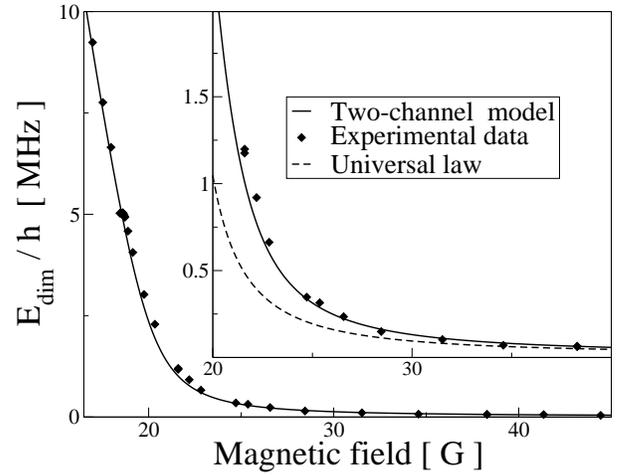}
}
\caption{
Comparison between experimental  dimer's energies in $^{133}$Cs 
obtained at Innsbruck (black losanges) 
\cite{Mar07} and results of the two-channel model for $b=0.7R_{\rm vdW}$ 
(black line). Inset: detail of the deviation of the spectrum 
from the universal law ${E_{\rm dim}=\hbar^2/(ma^2)}$ (dashed line).
}
\label{fig:Dim_Cs_12}
\end{figure}

We now determine the different parameters of the model from measured 
two-body properties. For this purpose we solve the two-body scattering 
problem in the center of mass  frame. For an incident plane wave of 
energy  ${E=\hbar^2 k_0^2/m}$ and momentum ${\mathbf k}_0$, the 
two-body state is a coherent superposition of one molecule plus two atoms: 
${|\Psi\rangle=(\beta b^\dagger_{\bf 0} + \int \dk  A_\kb a^\dagger_{\kb} a^\dagger_{-\kb} ) |0 \rangle}$, 
where $A_\kb$ is the atomic wave function in the form:
\begin{equation}
A_\kb = (2\pi)^3 \delta(\kb - {\mathbf k}_0) 
+ \frac{4 \pi \hbar^2}{m} \frac{f(E)}{2 \epsilon_{\mathbf k} -E -i 0^+} ,
\label{eq:A2b} 
\end{equation}
and $f(E)$ is the scattering amplitude. Remarkably, the $s$-wave 
scattering length (denoted by $a$) obtained in this two-channel model
can be \emph{exactly} identified  with the following expression 
which is known to be very accurate in the vicinity of a Feshbach resonance 
\cite{Julienne}: 
\begin{equation}
a = a_{\rm bg} \left( 1 - \frac{\Delta {\mathcal B}}{{\mathcal B}-{\mathcal B}_0} \right) .
\label{eq:a}
\end{equation}
In Eq.~\eqref{eq:a},  ${a_{\rm bg}}$ is the background  scattering 
length,  and ${\Delta {\mathcal B}}$ is the width of the resonance located at
the magnetic field  ${\mathcal{B}_0}$. One finds ${a_{\rm bg}=bg_0\sqrt{\pi}/(g_0-g_0^{\rm c})}$, with 
${g_0^{\rm c} = - 4\pi^{3/2}\hbar^2 b/m}$, an energy detuning from resonance 
${\nu = E_{\rm mol} - 2\Lambda^2/g_0  +{\delta \mu \Delta
{\mathcal B}}}$ where ${\nu = \delta \mu ({\mathcal B}-{\mathcal B}_0)}$  
and a resonance width 
${\Delta {\mathcal B} = 8 \pi \hbar^2 \Lambda^2 a_{\rm bg}/(m g_0^2 \delta \mu)}$. 
This last relation implies that the energy width ${\delta \mu \Delta {\mathcal B}}$ 
always has the same sign as the background scattering length $a_{\rm bg}$.  
We checked that this property is indeed verified for the various resonances reported 
in Ref.~\cite{Julienne}. Using these parameters, the scattering amplitude admits a 
simple expression for $E=-\hbar^2 q^2/m<0$ (${k_0 = iq}$ and ${q > 0}$):
\begin{equation} 
\frac{1}{f(E)}= q \mbox{erfc}(q b) - \frac{e^{-q^2 b^2}}{a_{\rm bg}} 
\left(1 - \frac{{\delta \mu \Delta {\mathcal B}}}{E - \nu  + {\delta \mu \Delta {\mathcal B}}} \right) .
\label{eq:scat-amplitude}
\end{equation}
We use experimental or theoretical spectroscopic data on the two-body bound 
states  as a way to choose a precise value for $b$ (in absence of such data,
we arbitrarily set $b=R_{\rm vdW}$). In order to avoid any confusion the 
two-body bound states are denoted by 'dimers'  and are distinct from the molecular 
state. Their binding energies ${E=-E_{\rm dim}= -\hbar^2 q_{\rm dim}^2/m<0}$ are
poles of ${f(E)}$. In what follows, we briefly sketch their spectrum as a function 
of the energy detuning $\nu$. For $\nu<0$, there is a branch terminating 
at zero energy for ${\nu =0^-}$: this branch results from the interchannel coupling 
and is denoted below as the Feshbach dimer's branch. Furthermore, in the case 
${a_{\rm bg} > b \sqrt{\pi}}$, another branch exists for all possible values of the
detuning. Away from the Feshbach resonance and for $\nu>0$, it results from 
the direct coupling in the open channel and we denote it as the 'background dimer' 
branch. For decreasing values of $\nu$ there is an avoided crossing between the 
two branches. We checked for several resonances that a choice of $b$ of the order 
of ${R_{\rm vdW}}$ permits to describe the lowest dimer's branch over a wide range 
of magnetic field detuning. In Fig.~\eqref{fig:Dim_Cs_12}, we compare the results 
of the model with the experimental data of the resonance located at ${{\mathcal B}_0 \sim -11.7}$~G 
for C{\ae}sium \cite{Mar07,Knoop} ($\Delta {\mathcal B} \simeq 28.7$~G, $a_{\rm bg} \simeq 1720 a_0$, 
$R_{\rm vdW} \simeq 101 a_0$ and $\delta \mu \simeq 2.3 |\mu_B|$ \cite{Julienne}). 
The system is in the vicinity of a shape resonance ${(a_{\rm bg} \gg R_{\rm vdW})}$ 
and the spectrum displayed corresponds to the branch of the background dimer. We 
found very good agreement for ${b=0.7R_{\rm vdW}}$.

We now turn to the central part of the present work by investigating the problem
of three interacting bosons. The general three-body eigenstate of Eq.~\eqref{eq:H} 
of energy $E$ in the total center of mass frame, is a coherent superposition of 
three atoms plus one atom and one molecule:
\begin{multline} 
|\Psi \rangle = \int \frac{d\Kb}{(2\pi)^3}\frac{d\kb}{(2\pi)^3} A_{\Kb,\kb} \,
a^\dagger_{\frac{\Kb}{2}+\kb} a^\dagger_{\frac{\Kb}{2}-\kb}  a^\dagger_{-\Kb} |0
\rangle \\+ \int \frac{d\Kb}{(2\pi)^3} \beta_\Kb b^\dagger_{\Kb}
a^\dagger_{-\Kb} |0 \rangle .
\label{eq:Psi3b}
\end{multline}
In Eq.\eqref{eq:Psi3b} $A_{\Kb,\kb}$ and $\beta_\Kb$ are respectively the atomic and the atom-molecule wavefunctions. For ${E>0}$, the atomic wavefunction contains an eigenstate of the atomic kinetic operator $A^{(0)}_{\Kb,\kb}$, and for all $E\in {\mathbb R}$, $A_{\Kb,\kb}$  can written as:
\begin{equation} 
\label{eq:A3b}
A_{\Kb,\kb} = A^{(0)}_{\Kb,\kb} + \Lambda \chi_\kb\beta^{\rm eff}_\Kb/
(E_{\rm rel} - 2 \epsilon_{\mathbf k}+i0^+) ,
\end{equation}
where $E_{\rm rel}=\hbar^2 k_{\rm rel}^2/m=E-3\hbar^2K^2/(4m)$ is the relative energy between a pair of atoms and the third atom. For convenience, we introduced in Eq.~\eqref{eq:A3b} the effective atom-molecule wavefunction $\beta^\mathrm{eff}_\Kb$:
\begin{equation}
\label{eq:beta_eff}
\beta^\mathrm{eff}_\Kb =  \beta_\Kb \times( E_{\rm rel} - \nu +{\delta \mu \Delta {\mathcal B}})/ 
(E_{\rm mol} - \nu + {\delta \mu \Delta {\mathcal B}}).
\end{equation}
The three-boson problem for this model is solved by finding the solutions of the integral 
equation which is deduced from the stationary Schr\"odinger equation: 
\begin{multline}
\frac{m |\chi_{\kb_{\rm rel}}|^2 \beta^{\rm eff}_\Kb }
{4 \pi \hbar^2 f(E_{\rm rel})} 
- 2 \int \dk  \frac{\beta^{\rm eff}_\kb \chi^*_{\frac{\Kb}{2}+\kb}
\chi_{\Kb+\frac{\kb}{2}}} {\epsilon_{\mathbf k} + \epsilon_{\mathbf K} 
+ \epsilon_{{\mathbf K} +{\mathbf k}} - E - i 0^+}\\
= - \int \dk \frac{\chi^*_\kb}{\Lambda} \left( A^{(0)}_{\Kb,\kb}
+ 2 A^{(0)}_{-\frac{\Kb}{2}+\kb , -\frac{3 \Kb}{4} -
\frac{\kb}{2} } \right).
\label{eq:3b}
\end{multline}
Remarkably all the two-body physics, contained in the scattering amplitude
$f(E)$, appears in the diagonal part of Eq.~\eqref{eq:3b}. Nearby a resonance, in the limit 
where ${|a| \to \infty}$, Eq.~\eqref{eq:3b} converges asymptotically in the low-energy regime (\emph{i.e.} for ${k b \ll 1}$ 
and ${K b \ll 1}$) toward the so-called Skorniakov Ter-Martirosian equation of Ref.~\cite{Sko57}, 
while the high energy limit of the integral kernel acts as an ultra-violet cut-off and the Thomas 
collapse is avoided \cite{Tho35}. As the magnetic detuning ${|\mathcal B - \mathcal B_0|}$ is increased, off-resonant 
effects come into play and the model can be used to quantify deviation from the universal 
theory \cite{Efimov,Braaten}. To this end, we numerically solved Eq.~\eqref{eq:3b} using two 
independent codes for several Feshbach resonances. We report in the following our results concerning 
the resonances studied experimentally in Refs.~\cite{Ketterle_Feshbach,Stenger,Kraemer,Knoop}.
\begin{figure}[h]
\resizebox{8 cm}{!}
{
\includegraphics{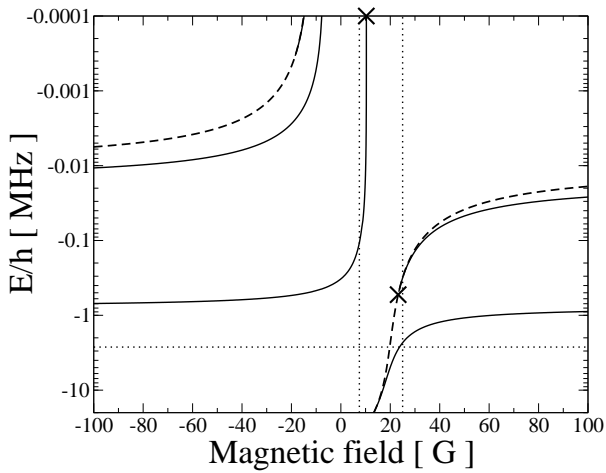}
}
\caption{
Spectrum of bound states in $^{133}$Cs (${{\mathcal B}_0 \sim -11.7}$~G). 
Solid lines: trimers; dashed line: dimers; dotted horizontal 
line: ``High energy'' limit ${-\hbar^2/(mR_{\rm vdW}^2)}$; 
doted vertical lines: position of the Efimov resonances 
observed in the experiment (at $7.5$~G and $25$~G resp.); Crosses: thresholds computed with 
the two-channel model (at $10.3$~G and $23.2$~G resp.).
}
\label{fig:Trim_Cs_12}
\end{figure}
First, we investigate the spectrum of trimers made of C{\ae}sium atoms for 
the Feshbach resonance at ${-11.7}$~G. Trimers are 
obtained by searching negative energy solutions (or ${E<-E_{\rm dim}}$ 
if a shallow dimer exists) of Eq.~\eqref{eq:3b} where ${A^{(0)}=0}$. 
At  resonance (${\nu=0}$), we recover the Efimov spectrum: ${E_n = E_0 e^{-2 \pi n/s_0}}$, 
$(n\in \mathbb N)$ where $E_0$ is the energy of the lowest Efimov state 
[of the order of $-\hbar^2/(m R_{\rm vdW}^2)$] and ${s_0=1.00624}$ 
\cite{Efimov,Braaten}. In Fig.~(\ref{fig:Trim_Cs_12}) we plot the 
spectrum as a function of the magnetic field. Our model predicts four 
important features: ${(i)}$ The existence of two Efimov branches  
extending in the regions of positive and negative ${\mathcal B}$.
Each branch is continuous through the formal limit ${|{\mathcal B}|=\infty}$. 
Other Efimov branches are located in a small interval near ${{\mathcal B}_0}$; 
${(ii})$ The threshold of the first Efimov branch at $10.3$~G 
observed in Ref.\cite{Kraemer} at $7.8$~G; ${(iii)}$ The two trimer branches 
hit the background dimer branch in a non universal region where short range details 
of the interatomic forces are not negligible --see inset of Fig.\eqref{fig:Dim_Cs_12}-- and where 
our model gives qualitative informations only; ${(iv)}$ In experiments reported in Ref.~\cite{Knoop}, 
the magnetic field was decreased from a large detuning and an atom-dimer
resonance loss was found at $\sim 25$~G. The present model shows that 
the observed threshold corresponds to the second trimer branch. The 
disagreement between theoretical and experimental thresholds follows from 
item ${(iii)}$. Moreover the asymptotic behavior of the wave function at 
distances ${r \ll a}$ is a crucial ingredient in Universal Theory 
and follows from taking the 'unitary approximation' of the scattering amplitude 
${f \sim -1/ik}$ at intermediate momentum ${1/a \ll k \ll \min (1/R_{\rm vdW},1/r_{\rm e}) }$,
where ${r_{\rm e}}$ is the effective range.  However, at  the observed threshold  
${\sim 25}$~G, ${a/R_{\rm vdW}\sim 4}$ and from Eq.~\eqref{eq:scat-amplitude},
${a/r_{\rm e}\sim 2}$. Hence, in this detuning region there is no clear 
separation of momentum scale and the 'unitary approximation' for 
intermediate momentum is not correct. This fact gives an important insight in the 
deviation from prediction of Ref.~\cite{Braaten} concerning the ratio between the scattering lengths at the 
two observed thresholds.

In this last part, we study the recombination of three incoming 
atoms of vanishing total energy into one two-body bound state 
and one atom. This is the main process responsible for the short 
life time of a resonant BEC in a dipolar trap. Experimentally, it is 
measured \emph{via} the atomic loss rate which is defined for 
$N$ atoms trapped in a cubic box of size $L$ by:
\begin{equation}
\dot{N} =  - \alpha_\mathrm{rec} N(N-1)(N-2)/L^6 ,
\label{eq:loss}
\end{equation} 
where  $\alpha_\mathrm{rec}$ is the three-body recombination constant. 
The atomic loss rate plotted as a function of the magnetic field 
detuning exhibits a large peak centered at the resonance. We limit 
our analysis to the standard regime where for a given detuning there 
exists at most one shallow  dimer only ($a_{\rm bg}$ is of the 
order of $R_{\rm vdW}$ or negative). There are two distinct regimes 
in the recombination process:  ${(a)}$ In the regime of negative energy 
detuning ($\nu<0$) the formation rate of Feshbach dimers is the 
dominant loss mechanism. This dimer is an eigenstate of the model 
Hamiltonian in Eq.~\eqref{eq:H} and  the recombination 
constant can be computed {\it exactly}; in what follows, it is denoted 
by ${\alpha_\mathrm{rec}^{\rm Fesh}}$; ${(b)}$ In the regime ${\nu>0}$ 
there is no shallow dimer and the deep bound states populated in inelastic 
scattering processes are not described by our model Hamiltonian. In this case,
we denote the recombination constant by ${\alpha_\mathrm{rec}^{\rm deep}}$.
In both regimes the source term in Eq.\eqref{eq:A3b} is ${A^{(0)}_{\Kb,\kb} 
=(2\pi)^6\delta(\Kb) \delta(\kb)}$ and corresponds to the incoming 
atomic wave function in the three-body scattering process. In the 
first regime [case ${(a)}$], the outgoing dimer and atom created by inelastic 
scattering have a relative momentum ${\frac{2}{\sqrt{3}} q_{\mathrm{dim}}}$.
The dimer formation manifests itself as an outgoing 
wave in the atom-molecule wave function, and the wave function $\beta_\Kb$ 
has a pole at ${K=\frac{2}{\sqrt{3}} q_{\mathrm{dim}}+i0^+}$; 
we denote the residue of this pole by $\gamma$. In order to 
evaluate ${\alpha_\mathrm{rec}^{\rm Fesh}}$, we enclose the three 
incoming atoms in a fictitious box of arbitrary large size $L$ and 
impose periodic boundary conditions. The wave function is then deduced 
from Eq.~\eqref{eq:Psi3b} with ${|\Psi^\mathrm{box} \rangle \simeq | \Psi \rangle / L^{9/2}}$.
The rate of molecule formation is obtained in configuration
space by computing the total flux of the probability current associated 
with the atom-molecule wave function (relative particle of reduced mass 
${2m/3}$) through the box surface. Since the flying dimer has a 
non vannishing probability ($p_\mathrm{closed}$) to be in the closed channel, we 
divide this total flux by $p_\mathrm{closed}$ thus obtaining the 
atomic loss rate $\dot{N}$. By comparing this expression with the 
case ${N=3}$ in Eq.~\eqref{eq:loss}, one obtains:
\begin{equation}
\alpha_\mathrm{rec}^{\rm Fesh} = 
2 \sqrt{3} \hbar q_{\rm dim}^3 |\gamma|^2 / (9 \pi m p_\mathrm{closed})  .
\label{eq:alpha_rec}
\end{equation}
\begin{figure}[t]
\resizebox{8 cm}{!}
{
\includegraphics{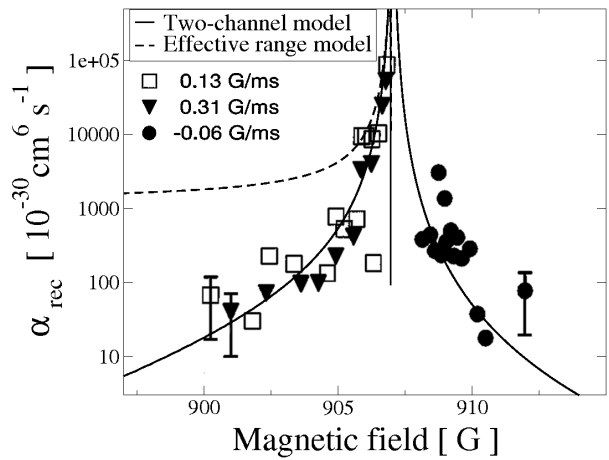}
}
\caption{Three-body recombination constant $\alpha_{\rm rec}$ as a 
function of the magnetic field $\mathcal{B}$ for the narrow Feshbach 
resonance in $^{23}$Na at 907~G. The three different symbols correspond 
to different ramp speeds of the magnetic field across the resonance in 
Ref.~\cite{Stenger}. Solid line: prediction of the two-channel model, 
dashed line: prediction of the effective range model in Ref.~\cite{Petrov3B}. 
The vertical line at resonance is due to the existence of narrow oscillations 
of ${\alpha_{\rm rec}}$ also predicted in Refs.~\cite{Petrov3B,Mora}.}
\label{fig:Rec_Na}
\end{figure}
In the regime ${\nu>0}$ [case ${(b)}$], the recombination rate can be 
evaluated on a qualitative basis only. The idea is to compute in the 
fictitious box of size $L$ the probability ${\mathcal P}_<$ of finding 
the three atoms or the atom and the molecule in a volume of the order of 
$R_{\rm vdW}^3$. This calculation is performed in configuration space by 
using Eq.~\eqref{eq:Psi3b}. Since $R_{\rm vdW}$ 
gives the typical size of deep bound states, the loss rate is obtained 
by dimensional analysis with ${\dot{N} \propto -\frac{\hbar{\mathcal P}_<}{mR_{\rm vdW}^2}}$, 
and finally the recombination constant is estimated by:
$
{\alpha_\mathrm{rec}^{\rm deep} = \frac{\hbar L^6{\mathcal P}_<}{mR_{\rm vdW}^2}}
$. We applied this formalism for the narrow resonance 
[\emph{i.e.} ${\delta \mu \Delta B \ll \frac{\hbar^2}{m a_{\rm bg}^2}}$] in 
$^{23}$Na at ${{\mathcal B}_0\simeq  907}$~G \cite{Ketterle_Feshbach} 
(${\Delta {\mathcal B} \simeq 1}$~G, ${\delta \mu \simeq 3.8 |\mu_B|}$, 
${a_{\rm bg} \simeq 63 a_0}$ and  ${R_{\rm vdW} \simeq 44.5 a_0}$ 
\cite{Julienne}). Fig.~(\ref{fig:Rec_Na}) shows a  dramatic agreement 
of our results (for $b=R_{\rm vdW}$) with the experiments in Ref.~\cite{Stenger}. 
Three-body properties in narrow resonances are usually  described within the 
effective range approximation by using the two parameters $a$ and  ${R^\star=\hbar^2/(m a_{\rm bg} \delta \mu \Delta B)}$ 
\cite{Petrov3B}. As shown in Fig.~\eqref{fig:Rec_Na}, 
this latter approach gives reasonable results very close to the resonance only. 
In the zero range limit ${(b\to 0)}$, when the parameters ${a}$ and ${R^\star}$ 
are held fixed, and ${{a_{\rm bg}} \equiv 0(b)}$ vanishes (but ${a_{\rm bg}\ne 
b \sqrt{\pi}}$),  the present model coincides with the effective range approach 
{\it exactly}. Therefore, in realistic situations where $b$ and $a_{\rm bg}$ are both 
finite, this two-channel model  quantifies consistently off-resonant effects. We verified 
that ${\alpha_{\rm rec}}$ in Fig.~\eqref{fig:Rec_Na} is not very sensible to the precise 
choice for ${b\equiv 0(R_{\rm vdW})}$, showing that for a narrow resonance the present
model gives quantitative results for the deviation from universality.

To conclude, we presented a rather simple formalism to capture the main physical 
features in resonant three-bosons systems. The short range details of interatomic 
forces are described by one parameter only, so that we avoided the 
complexity of more detailed models~\cite{Lee,D'Incao}.


We thank Y. Castin, F. Ferlaino, R. Grimm, F. Werner for fruitful 
discussions and W. Ketterle for providing us with experimental data. LPTMC is 
UMR 7600 of CNRS and its Cold Atoms group is associated with IFRAF.

\end{document}